\documentclass[12pt]{article}
 \usepackage{amsmath,amsfonts,latexsym,amssymb,amsbsy}
 \usepackage{times}
 \usepackage{color}
 \usepackage{graphicx}  

\DeclareMathAlphabet{\mathpzc}{OT1}{pzc}{m}{it}

\newcommand{\CC}{{\mathbb C}}
\newcommand{\RR}{{\mathbb R}}

\newcommand{\bcx}{\pmb{x}} 
\newcommand{\bcy}{\pmb{y}}

\newcommand{\bsfx}{\pmb{\mathsf{x}}} 
\newcommand{\sft}{\mathsf{t}}
\newcommand{\sfx}{\mathsf{x}}
\newcommand{\sfy}{\mathsf{y}}

\newcommand{\Hc}{{\cal H}} 
\newcommand{\Lc}{{\cal L}}
\newcommand{\Pc}{{\cal P}}

\newcommand{\pos}{{\mathpzc p}} 

\newcommand{\ie}{{\it i.e.\ }}
\newcommand{\etc}{{\it etc}}

\begin{document}
\title{{ \LARGE 
\hspace*{-27pt} Unruh versus Tolman: On the heat of acceleration
\hspace*{-27pt}} \\[0.5mm]
{\large Dedicated to the memory of Rudolf Haag}}

\author{\Large Detlev Buchholz${}^{(1)}$ and Rainer Verch${}^{(2)}$}

\date{
\small
${}^{(1)}$ Institute for Theoretical Physics, 
University of G\"ottingen, 37077 G\"ottingen, Germany\\[3pt]
${}^{(2)}$ Institute for Theoretical Physics, 
University of Leipzig, 04009 Leipzig, Germany}

\maketitle


\begin{abstract}
\noindent It is shown that the Unruh effect, i.e.\ the increase in temperature  
indicated by a uniformly accelerated thermometer in an inertial 
vacuum state of a 
quantum field, cannot be interpreted as the result of an exchange 
of heat with a 
{surrounding gas}. 
Since the vacuum 
is spatially homogeneous in the accelerated system its  
temperature must be zero everywhere as a consequence of Tolman's law. 
In fact, the increase of temperature of accelerated thermometers is due to 
systematic quantum effects induced by the local coupling between the 
thermometer and the vacuum. This coupling inevitably creates 
{excitations of}
the vacuum which transfer energy to the thermometer, gained by 
the acceleration, and thereby affect its readings. The  
temperature of the vacuum, however, remains to be zero for arbitrary 
accelerations.  
\\[10pt]
{\small Keywords: Unruh effect, Tolman-Ehrenfest law, Vacuum temperature }

\end{abstract}

\section{Introduction} 

In a well-known paper, Unruh \cite{Unruh:1976} considered an idealized, 
pointlike detector which follows a worldline
of constant proper acceleration in Minkowski spacetime while its 
degrees of freedom are coupled to a quantum 
field in the inertial vacuum state. He has shown that in the limit of 
large measuring times and of weak 
couplings the detector state will be found in a Gibbs ensemble 
corresponding to a temperature
which is proportional to the detector's proper acceleration; 
see also \cite{Sewell,Wa,dB-M,CrisHigMat}. 
The relation is 
\begin{equation} \label{eqn:Unruh-temp}
 T_D = \frac{a}{2\pi}
\end{equation}
where $T_D$ is the Gibbs ensemble temperature of the detector, $a$ is 
the proper acceleration along its worldline, and we use units 
where the velocity of light, Planck's constant and 
Boltzmann's constant have the values $c = \hbar = k = 1$.

Constant acceleration can be described as the effect of a constant 
gravitational field (by the equivalence principle) which in turn can 
be described by a static spacetime metric. In the case at hand this
is Rindler space whose metric is given in appropriate coordinates 
(assuming that the acceleration points into the $1$--direction) by  
\begin{equation}
ds^2 = (ax_1)^2 \, dt^2 -  d{\bcx}^2  \, , \quad  {\bcx} = (x_1,x_2,x_3) \, .
\end{equation}
The orbit of the thermometer is given in these (Rindler) coordinates 
by $x_1 = 1/a$, $x_2=x_3=0$ and $t$ denotes its proper time. 

 {It has been suggested to interpret the temperature $T_D$ 
of the detector as the temperature of a  relativistic  
gas which appears in the vacuum because of 
the acceleration, cf.\ \cite[p.\ 167]{FuRu}, 
\cite[p.\ 3721]{Unruh:1992}, \cite[p.\ 115]{Wa}}. 
If this picture is correct, \ie if one 
is effectively dealing in the accelerated vacuum with an 
equilibrium state of a gravitating gas, one can apply a  
classical result of 
 {Tolman \cite{Tolman:1930}  
and Tolman--Ehrenfest \cite{TolmanEhrenf:1930}}
who observed that the temperature in such systems is spatially 
varying. In the case 
at hand they obtain for the temperature $T({\bcx})$ 
at point ${\bcx}$ in Rindler space the relation 
\begin{equation} \label{Tolman}
T({\bcx}) \, a x_1 =   {\rm const}.
\end{equation}
where the constant depends on the system. 

If the Unruh 
temperature $\, T_D$ at $\, x_1 = 1/a$ is identified with the temperature 
of a gravitating  gas one obtains \ 
${\rm const.} = a / 2 \pi$, so the temperature 
of the vacuum  
depends on the position in the comoving system 
according to $T({\bcx}) = 1 / 2 \pi x_1$. Hence it 
is strongly varying with ${\bcx}$ and one would   
expect that the gas should also exhibit corresponding 
pressure and density variations. 

We will show in this article that this conclusion leads to 
contradictions, so there is no such gas. In particular, the 
(macroscopic) Tolman temperature and the (microscopic) Unruh 
temperature cannot be identified. This applies not 
only to the vacuum, but in fact to any accelerated 
equilibrium (KMS) state of a quantum field with respect to the 
time coordinate $t$. We will indicate the origin of 
this discrepancy, explain why Unruh detectors do not
describe perfect local thermometers and outline how local  
temperatures can be determined otherwise, leading to results
which consistently unify the Unruh effect and Tolman's law. Most 
arguments rest on results of our recent work \cite{DBRV:2014}. 

These results appear to have consequences also for other discussions
in which relation \eqref{Tolman} is of relevance, such as considerations 
of black branes \cite{RussoTownsend2010}, or the idea of a fundamental 
link between time and temperature (``thermal time'') 
\cite{HaggardRovelli2013}. At any rate, one has to be cautious
when identifying the Tolman temperature of~\eqref{Tolman} with 
temperature readings distorted by quantum effects, such as the 
Unruh temperature of an accelerated detector. 

\section{The Unruh detector, encore}
Turning to the details, let $\phi(\sfx)$ be a real, scalar quantum field 
on Minkowski space $\RR^4$, where we use inertial coordinates 
$\sfx = (\sfx_0, \sfx_1, \sfx_2, \sfx_3)$ (sans-serif letters). 
The action of spacetime translations $\sfy \in \RR^4$ 
and Lorentz transformations $\Lambda \in \Lc^\uparrow_+$ 
on the field is given by the map 
$\phi(\sfx) \mapsto \phi(\Lambda \sfx + \sfy)$ and the
field is assumed to be local, \ie $[\phi(\sfx), \phi(\sfy)] = 0$ if
$\sfx,\sfy$ are spacelike separated. 

We will consider different Hilbert space realizations 
(representations  \cite{Ha}) 
of this field which correspond to globally 
differing states. The 
basic reference state is the inertial vacuum, simply called vacuum
in the following, which is described by a unit 
vector $\Omega_0$ in the vacuum Hilbert space~$\Hc_0$. 
On~$\Hc_0$ there exists a continuous unitary 
\mbox{representation}~$U_0$ of the Poincar\'e group 
\mbox{$\Pc^\uparrow_+ = \RR^4 \rtimes \Lc^\uparrow_+$} 
such that (i) $\Omega_0$ is invariant under its action, (ii) the
generator (Hamiltonian) $P_0$ 
of the inertial time translations is positive and (iii)  
$U_0(\sfy,\Lambda) \phi(\sfx) U_0(\sfy,\Lambda)^{-1} = \phi(\Lambda\sfx + \sfy)$.
For the sake of concreteness, we take as a simple example fitting into 
this setting the theory of a free field of mass $m = 0$, acting on Fock 
space. But 
our arguments are, to a large extent, model independent.  

Since we are interested in the spatial dependence of temperature
in accelerated systems we consider a
Minkowski space based observer who enters with his clock a laboratory 
which, at some instant of time, is at rest and then undergoes constant 
acceleration $a > 0$ into the 1--direction. The laboratory is
assumed to have rigid walls which can withstand the tidal forces
caused by the acceleration, cf.\ \cite{DBRV:2014}. Using Rindler 
coordinates, the laboratory occupies at 
proper time $t \geq 0$ of the observer some region in the half--space
$L_t = \{ \bcx \in \RR^3 :  x_1 > 0 \}$ 
where the observer stays at $\bcx_o = (1/a, 0 , 0)$. 
Proceeding to Minkowski coordinates, this half--space corresponds to the half--hyperplane 
$L_t = \{ \sfx \in \RR^4 : 
\sfx_0 = \mbox{th}(at) \sfx_1, \sfx_1 > 0 \} = \Lambda_1(at) L_0$, 
where $\Lambda_1$ denotes the one--parameter group of Lorentz boosts 
into the 1--direction, parametrized by $at$.
Thus the time evolution of all points in the laboratory
is \mbox{determined} by this action.   

The observables carried along by the observer, testing the 
properties of the field are, at 
time \mbox{$t=0$}, described by polynomials 
$A = \sum c_n \phi(f_1) ... \pi(f_n)$ of the  
field operators $\phi$ and their canonical conjugates $\pi$  
which are integrated with test functions $f$ having  
support in the region $L_0$. 
Taking into account the preceding remarks about the evolution 
of points in $L_0$ and the transformation properties of the 
field under Lorentz transformations, it follows that the  
observables at time $t$ are given by (Heisenberg picture) \ 
$A(t) = U_a(t) A U_a(t)^{-1}$, 
where we have put $U_a(t) \doteq U_0(0,\Lambda_1(at))$.

Since the vacuum vector $\Omega_0$ is 
invariant under Lorentz transformations, 
the accelerated observer finds with his observables 
$A,B$ that the 
vacuum state $\omega_0$ is stationary,  \ie  
$\omega_0(A(t)) = \langle \Omega_0, U_a(t) A U_a(t)^{-1} \Omega_0 \rangle
= \omega_0(A)$. Moreover, as observed by 
Unruh \cite{Unruh:1976} and independently
by Bisognano and Wichmann \cite{BiWi}  (see also \cite{Fulling73}), 
the correlation functions $t \mapsto \omega_0(B A(t))$ 
satisfy the KMS condition which is a distinctive feature 
of equilibrium states \cite{HaHuWi,PuWo}. This
condition 
{can be presented in the form} \ 
$\omega_0(B \widetilde{A}(k)) = 
e^{k/T_D} \omega_0(\widetilde{A}(k) B)$, $k \in \RR$, 
where the tilde denotes the Fourier transform 
(in the sense of distributions) of the operator functions  
$t \mapsto A(t)$ and $T_D$ is the Unruh temperature
given above. Thus there arises the question of the physical 
significance of the parameter~$T_D$.

In order to answer this question, Unruh studied 
in \cite{Unruh:1976} the effect of the coupling of the accelerated  
vacuum state with a small system (probe). 
The simplest such example is a two
level system with Hilbert space $\CC^2$ and internal Hamiltonian
\mbox{$H_o = E_o \, \sigma_3$}, where $\sigma_0, \sigma_1, \sigma_2, \sigma_3$
are the Pauli matrices and $E_o$ is the internal energy 
relative to the time scale of the observer. The generator
of the time translations of the field in the 
accelerated laboratory is $a K_1$, where 
$K_1$ is the generator on $\Hc_0$ of the boosts in the 1--direction.
In order to describe the coupling between the
vacuum and the probe on the product Hilbert space
$\Hc_0 \otimes \CC^2$ we choose some smooth non--negative
function $\bcx \mapsto \pos(\bcx)$ that integrates 
to $1$ and has support in~$L_0$ about the chosen position of the probe
 {which may be distant from the position of the observer}.
Taking into account the redshift factor $ax_1$, which scales  
energies at the points $\bcx$ in the accelerated laboratory, the generator
of time translations of the coupled system is modeled by
\begin{equation} \label{generator}
G_{g, \pos} \doteq  aK_1 \otimes 1 + 1 \otimes E_{\pos} \, \sigma_3 
+ g \, \phi({\pos}) \otimes \sigma_1 \, ,
\end{equation}
where  
$E_{\pos} = \int \! d \bcx \, {\pos}(\bcx) ax_1 E_o$
is the internal energy of the probe at its respective position, 
$g$ is a coupling constant and $ \phi({\pos})$
is the field integrated with the function ${\pos}$. 
Thus the time translations of the coupled system relative to the 
proper time of the observer are described by the unitaries
$V_{g, \pos} (t) = e^{it G_{g, \pos}}$. 

Now let 
$\Omega_\otimes \doteq \Omega_0 \otimes \eta \in \Hc_0 \otimes \CC^2$
be the product of the vacuum vector and any given 
state vector $\eta$ of the probe and let 
$A_\otimes = \sum_i A_i \otimes \sigma_i $ be the observables of the coupled 
system. It has been shown by de Bi\'evre and Merkli \cite{dB-M} that 
for arbitrary coupling functions ${\pos}$
the expectation values of observables in the coupled state exist 
at large times, 
\begin{equation} \label{timelimit}
\lim_{t \rightarrow \infty} \langle \Omega_\otimes ,
V_{g, \pos}(t) A_\otimes V_{g, \pos}(t)^{-1} \Omega_\otimes \rangle 
\doteq \omega_{g, \pos}( A_\otimes) \, .
\end{equation}
These limits define stationary KMS states $\omega_{g, \pos}$ 
for the coupled dynamics corresponding to the same KMS parameter 
$T_D$ as for the uncoupled vacuum. Moreover, if one 
proceeds to small couplings one obtains 
\begin{equation}  \label{couplinglimit}
\lim_{g \rightarrow 0} \omega_{g, \pos}(A_\otimes)
= \sum_i \, \omega_0(A_i) \ \mbox{Tr} \big( (1/Z) e^{-E_{\pos} \sigma_3 / T_D} \, 
\sigma_i \big) \, .
\end{equation}
Even though the proof is rather   
involved, this result is physically not so surprising. For it
says that a microscopic probe cannot disturb an infinite 
equilibrium state, whilst it is itself driven to equilibrium,
described by a Gibbs ensemble.

 {In a similar manner one can treat the case of several probes, 
placed at different positions in the laboratory. There the internal 
Hamiltonian of the probe, appearing in the resulting Gibbs ensemble  
\eqref{couplinglimit}, has to be replaced by the sum of the 
respective internal probe Hamiltonians.}

This observation 
has been taken as justification to interpret probes as 
thermometers and to relate the KMS parameter $T_D$ of their 
ensembles to the 
temperature of the vacuum in the accelerated system, cf.\  
\cite{Unruh:1976,Sewell,dB-M}. As is apparent from the preceding
relation, $T_D$ does not depend on the position of the probes 
within the laboratory, fixed by the support of~${\pos}$.
This is in accordance with the known
fact that the KMS parameter of infinite equilibrium states 
is a global, superselected quantity~\cite{Ta}.
But it shows that probes cannot be used offhandedly to determine  
the temperature at their respective position. In 
fact, as has been explained, 
the local temperature of equilibrium states in the accelerated 
laboratory varies according to Tolman's law \eqref{Tolman}. 

Commonly, one copes with this problem 
by a reinterpretation of the 
readings of probes. Thinking of some fixed hardware, one 
compares the properties of the probe in the
accelerated system with those in an inertial equilibrium 
state~{\cite{Unruh:1976}}. 
By a rearrangement of the redshift factors,  
appearing in \eqref{couplinglimit} in the energy  
$E_\pos$, one argues that 
each probe behaves at its respective position in the 
laboratory as if it were exposed to an 
inertial equilibrium state of temperature 
$T_D / \int \! d \bcx \, {\pos}(\bcx) \, ax_1$. Thus
one \textit{defines} as ``true local temperature'' of the vacuum 
at point $\bcx$ 
the quantity $T_D(\bcx) = T_D / ax_1$, {seemingly} in accordance  
with Tolman's law. However, there then appears another conceptual 
problem.

\section{The vacuum seen from an accelerated laboratory}

The observer can determine, besides the temperature, other 
intensive properties of the vacuum,  
such as densities and pressures at different points 
in the laboratory.
These observables are described by operators~$A$ of the form  
given above. 
Note that the interpretation of these observables does not depend on the motion
of the observer, he can rely on their readings independently of
the dynamics.
A spatial shift $\bcy$   
and evolution in time $t$ of the observable $A$
results in the corresponding operator
$A(t,\bcy) = U_a(t) U_0(\bcy,1) A \, U_0(\bcy,1)^{-1} U_a(t)^{-1}$,
where we have identified Rindler and Minkowski coordinates
at time $t = 0$.   
Because of the invariance of the vacuum under Poincar\'e 
transformations, all expectation values of observables 
(hence also their variances \etc) 
satisfy  
\begin{equation}
\omega_0(A(t,\bcy)) =  
 \langle \Omega_0,  U_a(t) U_0(\bcy,1) A \, U_0(\bcy,1)^{-1} U_a(t)^{-1} \Omega_0 
\rangle =  \ \omega_0(A) \, .  
\end{equation}
Thus the vacuum is homogeneous also in the 
accelerated system, there appear no non--zero density or pressure
gradients. This fact is incompatible with a locally varying  
temperature of the vacuum state, in conflict with the above 
\textit{ad hoc} definition. Thus there is no hot gas appearing in  
accelerated vacuum states. 

This raises the question as to why the probe gains
energy from the vacuum in the accelerated system. 
 {As has been pointed out long ago,
cf.\ \cite[pp. 54-57]{BirrellDavies} and~\cite{CandelasDeutschSciama}, 
the answer rests upon 
the quantum nature of the coupling between the probe and the field,
given by the last term in relation~\eqref{generator}.} 
The operator $\phi(\pos)$ appearing there 
 {describes a local operation
in the region fixed by $\pos$ whose quantum effects 
inevitably change the energy content of the underlying ensemble,}
excitations are randomly created.  

Upon maintaining the acceleration of the laboratory, these 
{excitations} 
gain energy from the external forces, 
and they deliver parts of this energy to the probe in the course of time. 
Thus, due to these quantum effects, the probe also exchanges  
mechanical energy with its environment, not only heat as
is expected from a perfect thermometer.
The increase of energy of the probe due to 
this effect leads to an increase of temperature in its readings. 
In inertial systems or for small accelerations  
this effect causes errors in the temperature readings    
which lie beyond any measuring accuracy, so they do not matter.
But for large accelerations this systematic effect becomes prominent 
and can no longer be neglected.  

The replacement of $\phi(\pos)$ by another operator, 
coupling the probe with the vacuum, does not cure this problem.
In fact, there does not exist any non-trivial operator  
$A$ that is localized in the 
region $L_0$ and does not change the
energy content of the vacuum; 
 {this is a consequence of the 
Reeh--Schlieder theorem~\cite{StreaterWightman} according to
which the only local operators preserving the vacuum are multiples
of the identity.} 
Moreover, if one replaces in relation~\eqref{generator}
the field~$\phi(\pos)$ by another local operator $A$ one 
obtains in the limit of large measuring times and small 
couplings always the same final state of the probe, given 
in \eqref{couplinglimit}, cf.\ \cite{DeJa}. 
So the KMS parameter~$T_D$ does not depend on the specific nature 
of the coupling. All probes exhibit the same 
systematic error and indicate at asymptotic times their own 
temperature, induced by the measuring process,  
instead of the temperature of the vacuum.  

So what is the local temperature of the accelerated vacuum state?
In order to answer this question it has been proposed 
\cite{BuSo} to exhibit sufficiently many local observables $A$ which 
are appropriate to determine 
intensive properties of equilibrium states and to rely on the
zeroth law of thermodynamics and the Gibbs phase rule according to 
which the temperature of equilibrium states is uniquely fixed by 
these data. 
 {Quantum fluctuations can be
suppressed by proceeding to large time limits, respectively averages, of
these  observables}. 

This idea has been applied in~\cite{DBRV:2014} 
to, both, inertial and accelerated observers. 
Denoting by $U_0(\sft)$ the time translations 
on $\Hc_0$ in the inertial
system, 
 {the expectation values of all observables $A$ in states of $\Hc_0$ 
attain sharp values at asymptotic times.
Their (weak) limits are given by}  
\begin{equation} \label{state-function}
\lim_{\sft \rightarrow  \infty} U_0(\sft) A U_0(\sft)^{-1} = \omega_0(A) \, 1 \, ,
\end{equation}
fixing all intensive parameters in this case.
Performing the analogous limits in the accelerated laboratory 
with the corresponding time translations $U_a(t)$ one obtains
\begin{equation} \label{a-state-function}
\lim_{t \rightarrow  \infty} U_a(t) A U_a(t)^{-1} = \omega_0(A) \, 1 \, ,
\end{equation} 
\ie the asymptotic 
expectation values of the intensive observables are not affected
by the acceleration. Since all intensive parameters of the 
accelerated vacuum coincide with those 
in the inertial system one may conclude that the temperatures 
coincide as well, \ie the accelerated vacuum has
temperature zero everywhere.

\section{Local temperature observables}

The consistency of this approach has been tested in 
\cite{DBRV:2014} for arbitrary inertial and accelerated equilibrium states.
Assuming for simplicity that for each temperature 
(KMS parameter) $T > 0$ there exists only 
a single equilibrium state  $\omega_T$ in the inertial system,
one finds  that on 
the corresponding thermal Hilbert spaces~$\Hc_T$ there holds 
the analogue of relation~\eqref{state-function}, where on the right
hand side the vacuum state $\omega_0$ has to be replaced by
$\omega_T$. The functions $T \mapsto \omega_T(A)$ are the
macroscopic equations of state for the intensive observables $A$
in the inertial system.
(If, for given $T$, there exist several equilibrium states, 
these functions also depend on chemical potentials,
the phase structure, \etc.) It is evident that the 
value of $T$ can be recovered from the collection of these 
data, \ie temperatures can be determined 
with the help of the localized observables.   

In the uniformly accelerated laboratory there likewise exist
for all KMS parameters $T_a > 0$  
equilibrium states $\omega_{T_a}$ with regard to the 
accelerated dynamics, given by the adjoint action of the 
unitaries $U_a(t)$ on the observables \cite{HaNaSt}. 
In order to simplify the
notation we adopt here the convention 
that all quantities with an index $a$ refer to 
this dynamics. In particular, the vacuum 
corresponds to the Unruh parameter $T_{a, 0} = a / 2 \pi$, i.e.
$\omega_{T_{a,0}} = \omega_0$ on all observables in the 
accelerated laboratory. Again, there holds an 
analogue of relation~\eqref{a-state-function}
on the Hilbert spaces~$\Hc_{T_a}$ attached to 
the accelerated equilibrium states, where one now has to 
replace $\omega_0 = \omega_{T_{a,0}}$ by 
$\omega_{T_a}$ for given KMS parameter $T_a$.  

In order to determine the thermal interpretation of the KMS 
parameters $T_a$ in the accelerated system one  
compares the expectation values of local observables $A$ 
in the accelerated equilibrium states with those in the inertial system.
This approach is analogous to that used for probes, where the  
temperature scale is likewise calibrated in inertial equilibrium states.  

In the present simple free field model one may take 
as a ``local thermometer'' \cite{BuSo,BuOjRo} 
the normal ordered square of the field, 
$\Theta^{} \doteq 12 : \! \phi^2 \! :$ 
(or any other of its even powers). The 
numerical factor is suggested by calibration in the inertial 
equilibrium states $\omega_T$, giving 
$\omega_T(\Theta^{} (\bsfx)) = T^2$ 
for any $T \geq 0$. So the readings of 
$\Theta^{}$ indicate the square of the
temperature which is equal 
at all points $\bsfx$ in the inertial system. Plugging this 
observable into the accelerated 
equilibrium states one obtains~\cite{BuSo} 
\begin{equation}
\omega_{T_a}(\Theta^{} (\bcx)) = 
\big(T_a^2 - (a/2 \pi)^2 \big)/(ax_1)^2 \, .
\end{equation}
Hence, for given KMS parameter $T_a \geq a/2 \pi$,  
the thermometer indicates at any given point $\bcx$ 
in the accelerated laboratory the temperature 
\begin{equation} \label{Tolman-correct}
T_a(\bcx) = \sqrt{(T_a^2 - (a/2 \pi)^2 \big)}/ax_1 \, .
\end{equation}
This relation is consistent with Tolman's law \eqref{Tolman} with 
\ $\mbox{const.} = \sqrt{(T_a^2 - (a/2 \pi)^2)}$. 
 {Notably, the systematic error in the readings  
of probes is corrected} 
and the result is in accord with the statement that the temperature is~$0$ 
everywhere for $T_a = a / 2 \pi$, \ie 
in the vacuum state. Note that 
Tolman's law in this concrete form 
is obtained here as a result, it is not put in by hand. 

It also follows from relation \eqref{Tolman-correct}
that the temperature tends to zero
in all accelerated equilibrium states at sufficiently large 
distances from the boundary of $L_0$.
As a matter of fact, in these remote regions 
the expectation values of all local \mbox{observables} 
in the accelerated KMS states coincide with those in 
the vacuum~\cite{DBRV:2014}. Thus the observer 
can calibrate his observables up there according to 
inertial standards. 

For KMS parameters $T_a < a/2 \pi$, the expectation values of $\Theta$ 
in the corresponding KMS states are negative; hence one cannot
assign to them a meaningful temperature. This can be understood 
if one notices that also all densities and pressures are negative in these 
states (taking the remote vacuum as a reference).
 {In the presence of acceleration, the 
excitations created by local measurements effectively 
equilibrate these states}, 
but from an inertial point of view they are to be regarded as 
ensembles which are far from (local) equilibrium. 
Note that the restriction of any accelerated KMS state 
to the observables in any given compact region can be 
represented by 
 {density matrices} 
in Fock space~\cite{VeSa}. 
Hence an accelerated observer,  launched in 
Minkowski space where he has calibrated his observables, can interpret 
in these terms the properties of the states and has    
no reason to rely on elusive Rindler quanta. 

So we conclude that the vacuum does not exhibit any 
thermal properties in accelerated laboratories, its temperature
remains to be zero. The increase in temperature indicated 
by microscopic probes is due to the quantum induced creation of 
excitations caused by the interaction; they transmit 
energy to the probes, gained from the accelerating 
forces. This energetic quantum effect can be understood in rough analogy 
to the production of heat by friction, 
but it is not the result of an exchange of thermal energy between
probes and a heat bath (Rindler gas). As we have shown here, 
the latter interpretation
is not tenable on several theoretical grounds. 

\vspace*{5mm} \noindent
{\bf \large Acknowledgements}
\\[6pt]
We acknowledge stimulating critical comments by S.A.~Fulling, G.L.~Sewell 
and W.G.~Unruh and an intense exchange with R.M.~Wald on our differing views.

\end{document}